\documentclass[conference]{IEEEtran}
\IEEEoverridecommandlockouts
\usepackage{cite}
\usepackage{amsmath,amssymb,amsfonts}
\usepackage{algorithmic}
\usepackage{graphicx}
\usepackage{textcomp}
\usepackage{xcolor}
\def\BibTeX{{\rm B\kern-.05em{\sc i\kern-.025em b}\kern-.08em
    T\kern-.1667em\lower.7ex\hbox{E}\kern-.125emX}}
\begin{document}

\title{DataAI-6G: A System Parameters Configurable Channel Dataset for AI-6G Research }
\author{Zibing Shen, Jianhua Zhang, Li Yu, Yuxiang Zhang, Zhen Zhang, Xidong Hu\\ State Key Laboratory of Networking and Switching Technology, Beijing University of Posts and Telecommunications,\\Beijing, China\\
Email: \{szb, jhzhang, li.yu, zhangyx, zhenzhang, hxd\}@bupt.edu.cn}

\maketitle

\begin{abstract}
With the acceleration of the commercialization of fifth generation (5G)
mobile communication technology and the research for 6G communication systems, 
the communication system has the characteristics of high frequency, multi-band, high speed movement of users and large antenna array.
These bring many difficulties to obtain accurate channel state information (CSI), which makes the performance of traditional communication methods be greatly restricted. Therefore, there has been a lot of interest in using artificial intelligence (AI) instead of traditional methods to improve performance. A common and accurate dataset is essential for the research of AI communication.
However, the common datasets nowadays still lack some important features, such as mobile features, spatial non-stationary features etc.
To address these issues, we give a dataset for future 6G communication. In this dataset, we address these issues with specific simulation methods and accompanying code processing.
\end{abstract}


\section{Introduction}
6G mobile networks are expected to support further 
enhanced mobile broadband, ultramassive machine-type, enhanced ultrareliable and low-latency, long-distance, and high-mobility communications and other 
emerging scenarios for the 2030 intelligent information society, which requires instantaneous, extremely high-speed wireless connectivity \cite{b1}.
These new scenarios and requirements make it necessary to consider an increasing number of features when modeling the channel. The channel model based on statistical characteristics becomes more and more complex as the number of characteristics considered increases, and an overly complex model is not conducive to future research. In order not to further complicate the model, researchers have come up with the idea of using AI techniques instead of, or in addition to, optimizing the traditional modeling approach. This idea is well supported in today's era of big data.
\par
With the continuous exploration of researchers, machine learning (ML)-
based AI techniques have become the
key to develop the next-generation communication system \cite{b2}.
High-mobility communications make the CSI tends to be 
out of date in a short time period, multi-antenna and multi-band make acquiring CSI difficult and requires significant overhead, and with the usage of ultramassive MIMO, the energy consumed by signal 
transmission and RF chains will become considerable. These make it very difficult to obtain channel information in the space, time, and frequency domains. In order to get accurate CSI and reduce overhead, AI-based time-, frequency-, and space- domain channel extrapolation \cite{b3} and compressive sensing for massive MIMO CSI feedback \cite{b4} have been presented.
In the millimeter wave band, blocking has a significant impact on the quality of communication and the overhead of beam selection is huge, which are the challenges of future high frequency communication. In \cite{b5,b6}, two AI-based methods for blockage prediction and beam prediction are proposed, and both of these methods effectively solve the above problems. In addition, the prediction of a particular channel characteristic, such as path loss \cite{b7}, can also be very useful to further improve the communication quality.
\begin{figure}[htbp]
	\centering
	\includegraphics[scale=0.24]{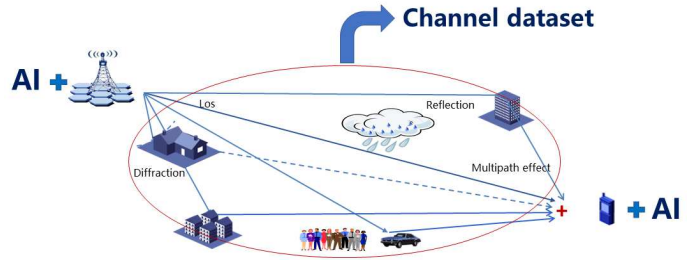}
	\caption{Communication with AI.}
	\label{figure1}
\end{figure}
\par
Adding AI at the base station and user side can improve the performance and reduce the overhead of communication. And to implement these AI applications, a large amount of channel data is necessary. The set of these data is the essential channel dataset in AI training, as shown in Fig. 1.
To meet the data requirements of the researcher, we have introduced the DataAI-6G dataset, which is designed for
machine learning research in wireless channel transmission and modeling. More specifically, using this dataset, researchers can easily construct the inputs and outputs of several machine learning applications. The DataAI-6G dataset provides angle of departure (AOD), angle of arrival (AOA), delay, phase, power of each path and the path loss between any pair of
transceiver antennas. These data are obtained from the
ray-tracing simulator, Wireless InSite, developed by Remcom \cite{b8}. Remcom Wireless InSite, is widely used in
mmWave and massive MIMO research at both industry
and academia, and has been verified with real-world channel measurements \cite{b9,b10,b11}. 
More important, our dataset has a dedicated set of codes, which can synthesize the UL and DL channel matrices and has user moving function. More details will be discussed in the rest of this paper. 

\section{Design of dataset}
According to \cite{b12}, the use of ultra-large antenna arrays introduces near-field spatial non-stationary features, which is an essential feature in 6G communications.
In \cite{b13}, the existence of UL to DL mapping has been proved, so there are many researchers are studying the mapping of UL to DL. In the future wireless communication, high-speed movement features are the focus of attention, but users in existing datasets are usually static.
So, in our dataset, we have considered the above three features. Firstly, in the data simulation stage, we simulate each antenna array element separately instead of using the plane wave synthesis method. Then in the the code synthesis stage, the dataset can synthesize the CSI of the UL/DL channel using the angle, delay, power, and phase information of each path. More importantly, the dataset is able to introduce further Doppler phase shifts on this basis to obtain the CSI of the moving state. The specific synthesis principle is as follows.
\par
In our dataset, multi-antenna technology has been considered.
In order to get the channel matrix, we need calculate the channel impulse response (CIR) of each antenna pair first. Consider a MIMO system with multiple base stations and multiple user areas. For the k-th antenna
at the base station x and the g-th antenna at the u-th user point
in the y-th user area, there will be a large number of Multipath
components (MPC) between them. So the CIR can be written as
\begin{equation}
	h_{x_k,y_{ug}}=\sum_{i=1}^{M}\alpha_ie^{j\varphi_i}\delta(\tau-\tau_i),
\end{equation}
where $\alpha_i$ and $\varphi_i$ represent the amplitude and the phase of the i-th path, respectively. $M$ denotes the total number of paths between these two antennas and $\tau_i$ denotes the delay of the i-th path.
\par
However, in real-world communication, the receiving antenna usually samples the received signal at a certain frequency, so the received signal will be divided into multiple time-delayed distinguishable paths. In the DataAI-6G dataset, we simulate this reception method to obtain the channel response that most closely resembles the actual situation.
Assuming that the receiver samples the received signal at a sampling interval of 1/BW (BW is the channel bandwidth), 
the channel impulse response at the i-th sampling interval can be expressed as
\begin{equation}
	h^i_{x_k,y_{ug}}=(\sum_{n=1}^{N_i}\alpha_ne^{j\varphi_n})\delta(\tau-\tau_i),
\end{equation}
where $\alpha_n$ and $\varphi_n$ represent the amplitude and the phase of the n-th path, respectively.
$N_i$ denotes the total number of paths in the i-th sampling interval and $\tau_i$ denotes the
delay of the i-th sampling interval.
\par
After that, the impulse responses of all sampling intervals are superimposed and
converted to the frequency domain to obtain the frequency domain channel response. 
Then, the user-set UL/DL carrier frequency is brought into the formula of frequency domain channel response to obtain a complex value, and this complex value is stored as an approximate channel response in the generated dataset.
The UL/DL channel response can be written as
\begin{equation}
	\begin{cases}
		H^{up}_{x_k,y_{ug}}=\sum_{i=1}^L(\sum_{n=1}^{N_i}\alpha_ne^{j\varphi_n})e^{-j2\pi{f_{up}}{\tau_l}},\\
		\\
		H^{down}_{x_k,y_{ug}}=\sum_{i=1}^L(\sum_{n=1}^{N_i}\alpha_ne^{j\varphi_n})e^{-j2\pi{f_{down}}{\tau_l}},
	\end{cases}
\end{equation}
where L denotes the number of sampling intervals and $f_{up}$/$f_{down}$ denotes the UL/DL carrier
frequency.
Due to the difference of UL and DL carrier frequencies, the UL and DL channel response will be different in magnitude and phase. And since the UL and DL channel responses are calculated using similar formulas, there is a strong correlation between them. The advantage of using this approach to obtain the UL and DL channel responses is that the researcher has the flexibility to set the UL and DL carrier frequencies. However, since only the simulation data of the DL channel are available, this synthesis can only approximate the UL channel, which is still lacking in terms of accuracy. To further improve the accuracy of the UL and DL channels, UL simulation data or actual measurement data can be included in future studies.
\par
On the basis of the UL/DL features, we will proceed to discuss the mobile features. To get the mobile features, we need to take Doppler phase shift into consider. The expression of the Doppler phase shift can be written as
\begin{equation}
	\Delta\varphi=2\pi\frac{\boldsymbol{\rm{v}}\cdot\boldsymbol{\rm{n}}}{\lambda_c}\Delta t,
\end{equation} 
where $\boldsymbol{\rm{v}}$ denotes the velocity vector in the direction of movement and $\boldsymbol{\rm{n}}$ denotes the direction vector of AOA in DL channel and negative direction vector of AOD in UL channel. $\lambda_c$ and $\Delta t$ represents the wavelength of the carrier wave and time interval, respectively.
\par
After get the Doppler phase shift, we add it to Eq.(3). For the k-th
antenna at the base station x and the g-th antenna at the u-th user point in the y-th user area,  the frequency
domain channel response in the moving state can be written as
\begin{equation}
	\begin{cases}
		H^{up}_{x_k,y_{ug}}=\sum_{i=1}^L(\sum_{n=1}^{N_i}\alpha_ne^{j(\varphi_n+\Delta\varphi)})e^{-j2\pi{f_{up}}{\tau_l}}.\\
		\\
		H^{down}_{x_k,y_{ug}}=\sum_{i=1}^L(\sum_{n=1}^{N_i}\alpha_ne^{j(\varphi_n+\Delta\varphi)})e^{-j2\pi{f_{down}}{\tau_l}}.
	\end{cases}
\end{equation}
The channel response obtained in this way contains mobile features, so our dataset is well suited to researchers for the study of mobile features.
\begin{figure*}[htbp]
	\centering
	\includegraphics[scale=0.126]{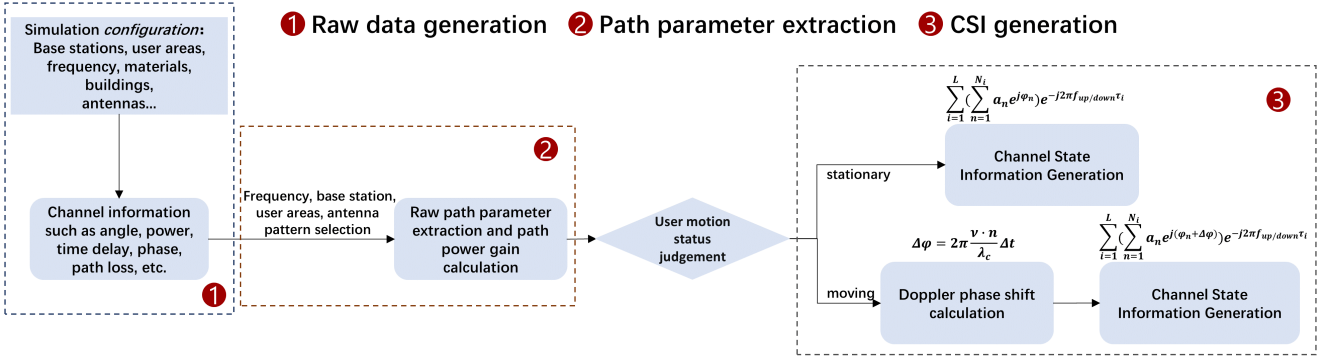}
	\caption{Framework of the dataset.}
	\label{figure2}
\end{figure*}
\begin{figure*}[htbp]
	\centering
	\includegraphics[scale=0.25]{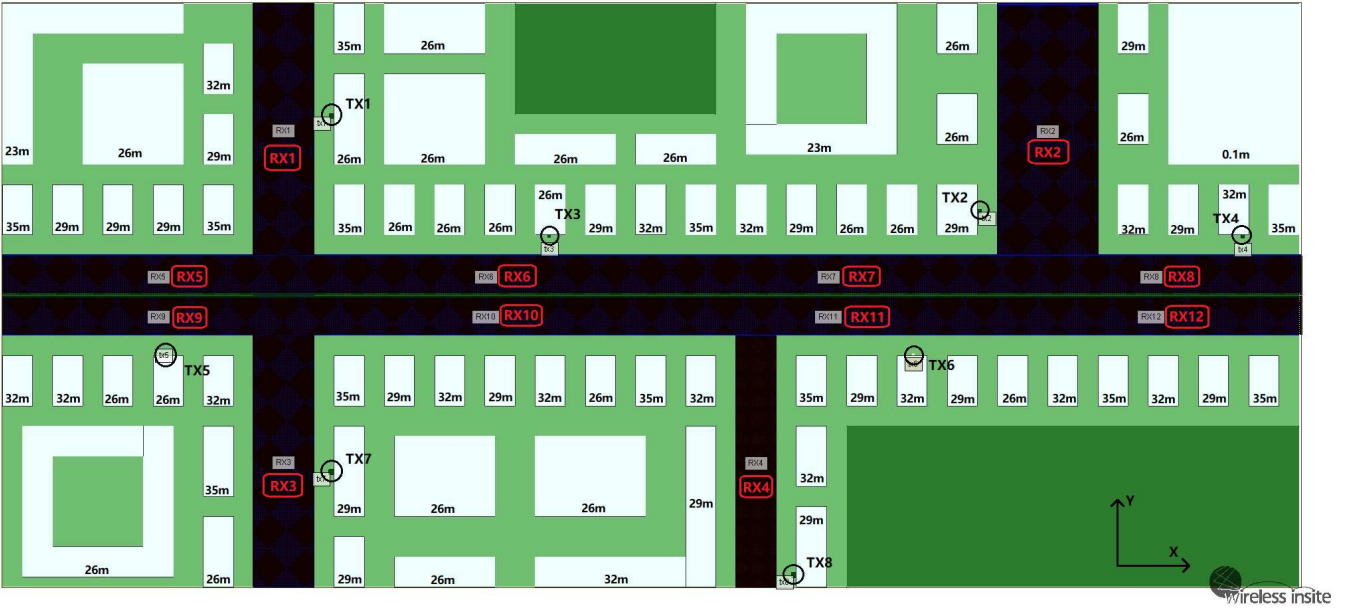}
	\caption{Outdoor street scenario.}
	\label{figure3}
\end{figure*}
\section{Dataset generation}
We build a large outdoor street scenario with multiple configurations in multiple bands
using Wireless Insite \cite{b8} and simulate it to obtain a set of channel parameters. We
provide a generic framework that allows researchers the flexibility to configure some
parameters in the code according to their needs. As shown in Fig. 2, the researchers can then bring the
raw channel parameters as input to the framework to output the customized dataset.

\subsection{Outdoor street scenario}
The outdoor street scenario is dedicated to providing researchers with diverse scene
features to meet the needs of machine learning different requirements. The whole
scenario is 646 m long and 290 m wide, which is an extensive outdoor scene, as shown
in Fig. 3. Two horizontally oriented main streets run through the whole scenario,
and four vertically oriented secondary streets are connected to the horizontally oriented
ones. To provide multi-regionalized data, we set up at least one base station for each
street. In total, we build 8 BS and 12 user grids, which are scattered within 6
streets. The users on the streets are evenly distributed within the grid. In addition, the streets are flanked by buildings of different heights and vegetation of varying sizes.
For simplicity, the buildings are rectangular and solid, so that the rays from the base
station cannot penetrate the buildings.
\par
More detail, the locations of these 8 BS are distributed on both sides of the street. Four of the base
stations are set up in two main streets in a horizontal direction, and four base stations
are respectively set up in four streets in a vertical direction. Each base station is
equipped with different types of antennas as well as different heights. TX2 and TX5
are equipped with a single element, which is the omnidirectional antenna, and the rest
of the base stations are equipped with multiple antennas. It is necessary to elaborate
that each array element constituting the MIMO antenna array is a half-wave dipole,
and the distance between them is half a wavelength. Users are evenly distributed among 12 user grids, and each starting point of the user grid is
located in the left corner. An example of the user points arrangement is shown in Fig. 4. The users in RX1, RX2 and RX3 are equipped with a 2$\times$2 uniform planar array, 
and the other users in the rest of the user gird are equipped with a omnidirectional antenna.
\begin{figure}[htbp]
	\centering
	\includegraphics[scale=0.06]{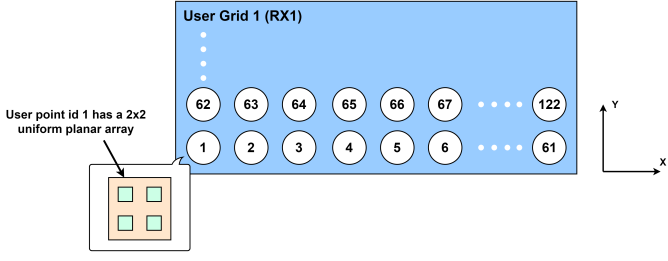}
	\caption{User points arrangement.}
	\label{figure4}
\end{figure}
\par
We use the X3D model, which is by far the most versatile, functional, and accurate
propagation model in Wireless Insite. Considering the meaningful received power, for
simplicity, only the first 4 reflections are considered. More importantly, the accuracy
of blocking and beam prediction can be further improved by exploiting the diffraction
properties \cite{b5}, \cite{b6}. But on the other hand, the received power decreases significantly
as the number of diffractions increases, so we turn on only one diffraction. After we configure the main parameters in Wireless Insite, it performs signal
propagation simulation and finally gives ray tracing results. The results of the simulation contain (i) the azimuth and elevation angles of departure of each path, (ii) the azimuth and elevation angles of arrival of each path, (iii) the path receive power, (iv) the path phase and (v) the propagation delay of each path. 
Wireless Insite can also output the overall received power, the overall phase, and the path loss of a receive point for all valid paths.
\subsection{Advantages of dataset}
Compared with other datasets, such as DeepMIMO \cite{b14},
Wireless AI Research Dataset \cite{b15}. Our dataset has three major
advantages(as shown in Table 1): (i) Spatial non-stationary features are considered in the simulation. (ii) With the user moving function that considers Doppler, users can freely
configure the moving route and moving speed. (iii) Has the ability to
generate any number of user points. In the next of this part, we will explain in detail how the last two functions are implemented in the code.
\begin{table*}[htbp]   
	\begin{center}   
		\caption{COMPARISON BETWEEN DATASETS}  
		\label{table:1} 
		\begin{tabular}{|c|c|c|c|}   
			\hline   \textbf{Dataset} & \textbf{DeepMIMO} & \textbf{Wireless AI Research Dataset} & \textbf{DataAI-6G}\\   
			\hline Multi-band & \checkmark & \checkmark & \checkmark  \\ 
			\hline Spatial non-stationary features &  &  & \checkmark  \\ 
			\hline Number of antenna configurations & \checkmark & \checkmark & \checkmark   \\  
			\hline Antenna rotation and pattern & \checkmark  & \checkmark  & \checkmark \\
			\hline Arbitrary multi-user point &   &   & \checkmark\\
			\hline Selective activation of BS and users & \checkmark & \checkmark & \checkmark  \\
			\hline User moving function with Doppler &   &   & \checkmark \\
			\hline Customized moving path and speed &   &   & \checkmark \\
			\hline BW Customization & \checkmark & \checkmark & \checkmark \\
			\hline  
		\end{tabular}   
	\end{center}   
\end{table*}
\par
In the profile of the code, the researcher can select the base station and the user area to be activated, and can also select the desired user points in the user area, while the number of antennas of users and base stations can be freely set according to the requirements. After setting the above parameters, researchers can choose the frequency of the channel(3.5 GHz, 28 GHz or 60 GHz), the antenna pattern of users and base stations, the bandwidth and carrier frequencies of UL/DL channel. Then the code will extract the AOA, AOD, power, delay and phase of each path, and the path loss of the channel will also be extracted.
After obtained the angle, phase, delay and power information of each path,
The dataset will synthesize the channel response using Eq.(3).
\par
If researchers want the user to move in the user grid, they just need set the parameter 'move' to 't'. In the DataAI-6G dataset, the user can move along four directions: up, down, left and right . The researcher only needs to set the corresponding parameters in the configuration file to specify both the path and direction of movement. Then, the code will perform point sampling on the movement path
according to the user-set speed and sampling interval. In order to calculate the Doppler phase shift due to the movement, the distance difference between the virtual point that get by point sampling and the user point in the user gird has to be calculated first.
\begin{equation}
	\Delta d=\kappa v\Delta t -m\Delta s,
\end{equation}
where $\kappa$ means the $\kappa$-th sample point, $\Delta s$ denotes the interval between real user points, $v$ and $\Delta t$ represent the speed of the user and the sampling interval. $m$ in Eq.(6) means the $m$-th user point in the moving path, which calculated by $\lceil\frac{\kappa v\Delta t}{\Delta s}-1\rceil$. Then, the code will bring $\Delta d$ into Eq.(4) to calculate the Doppler phase shift, which can be written as 
\begin{equation}
	\Delta \varphi=2\pi\Delta\rm{d}\frac{\boldsymbol{m}\cdot\boldsymbol{\rm{n}}}{\lambda_c},
\end{equation}
where $\boldsymbol{m}$ denotes the unit vector in the direction of movement and $\boldsymbol{\rm{n}}$ denotes the direction vector of AOA/AOD in DL/UL channel. $\lambda_c$ represents the wavelength of the carrier wave. After get the Doppler phase shift, the dataset will synthesize the channel response in the moving state using Eq.(5).

\section{Case of beam prediction}
In this section, we will use a beam prediction algorithm to validate our dataset.
\par
We consider a mobile cellular network including one base station (BS) and one moving user equipment (UE). The UE is communicating with the BS, and both line-of-sight (LOS) and none-line-of-sight (NLOS) exist during the movement. Since the future networks are likely to coexist in sub-6 GHz and mmWave bands, we assume that the BS is equipped with two antenna arrays. One works at the sub-6 GHz band and the other works at the mmWave band. 
\par
In our method, only the sub-6 GHz uplink (UL) channels is utilized for
beam prediction. During the UL signaling, UE sends pilot 
signals to the BS in each scheduling time frame, and the BS
receives the UL signal. Denote $\boldsymbol{\rm{y}}_{up}[k]$ as the received UL											
signal at the k-th subcarrier, $\boldsymbol{\rm{y}}_{up}[k]$ can be shown as
\begin{equation}
	\boldsymbol{\rm{y}}_{up}[k]=\boldsymbol{\rm{h}}_{up}[k]s_p+\boldsymbol{\rm{n}}_{up}[k],
\end{equation}
where $\boldsymbol{\rm{h}}_{up}[k]$ denotes the UL sub-6 GHz channel and $s_p$
denotes the signal transmitted from UE. $\boldsymbol{\rm{n}}_{up}[k]$ represents
the additive white Gaussian noise (AWGN).
\par
Let $\boldsymbol{\rm{h}}_{down}[\overline{k}]$ denotes the DL channel. The received signal of the
UE at both sub-6 GHz and mmWave bands is given by
\begin{equation}
	\boldsymbol{\rm{y}}_{down}[\overline{k}]=\boldsymbol{\rm{h}}_{down}[\overline{k}]\boldsymbol{\rm{f}}s_d+\boldsymbol{\rm{n}}_{down}[\overline{k}],
\end{equation}
where $s_d$ represents the signal transmitted from the BS and $\boldsymbol{\rm{n}}_{down}[\overline{k}]$
represents the AWGN. For the sub-6 GHz band, $\boldsymbol{\rm{f}}$ denotes the
sub-6 GH beamforming (BF) vectors which can be
obtained by matched filtering. $\boldsymbol{\rm{f}}_{sub6}$ can be written as
\begin{equation}
	\boldsymbol{\rm{f}}_{sub6}=\frac{\boldsymbol{\rm{h}}^*_{up}[k]}{\left| \boldsymbol{\rm{h}}_{up}[k] \right|}.
\end{equation}
\par
In the millimeter wave band, a large number of antennas will be used, resulting in a high overhead using the direct calculation method. Therefore, in order to reduce the overhead in high-band communications, we generally use codebooks for beam selection,
$\boldsymbol{\rm{f}}_{mmW}\in F_{mmW}$ which denotes
the mmWave BF vectors. $F_{mmW}$ is a set of pre-prepared
beamforming vectors. Denote $\frac{P}{\sigma^2}$ as the DL transmit signal-to-noise ratio (SNR).
The DL data rate for both sub-6 GHz and mmWave channels
can be shown as
\begin{equation}
	R(\boldsymbol{\rm{h}}_{down}[\overline{k}],\boldsymbol{\rm{f}})=Blog_2(1+\frac{P}{\sigma^2}\left| \boldsymbol{\rm{h}}_{down}[\overline{k}]\boldsymbol{\rm{f}} \right|^2).
\end{equation}
The optimal mmWave BF vector $\boldsymbol{\rm{f}}^*$ is selected to maximize
the mmWave rates $R$. And the optimal BF vector $\boldsymbol{\rm{f}}^*$ is
utilized to train the machine learning model. $\boldsymbol{\rm{f}}^*$ can be given by
\begin{equation}
	\boldsymbol{\rm{f}}^*=\mathop{argmax}_{\boldsymbol{\rm{f}}_{mmW}\in F_{mmW}}R(\boldsymbol{\rm{h}}_{down}[\overline{k}],\boldsymbol{\rm{f}}_{mmW}).
\end{equation}
\par
In this method, we will use the model in \cite{b5} and the DataAI-6G dataset to predict the DL optimal beam at 60 GHz at time slot t+1 using the UL channel response at 3.5 GHz from time slot t-24 to time slot t. To select the optimal beam of 60 GHz for
training and testing, an N-phase codebook $C$ is utilized. Each
code in $C$ can be utilized to generate a beam $\boldsymbol{\rm{f}}_{mmW}$, and all
beams form a beam set $F_{mmW}$. The method of selecting the optimal beam from $F_{mmW}$  is shown in Eq.(12).
\par
We choose TX3 BS and RX6 UE area, and set the user moving in this area at the speed of 72 km/h, 90 km/h and 108 km/h. The user moves in the positive direction of the x-axis with the sample frequency of 1 kHz. In order to be able to compare with the dataset in \cite{b5}, we take more than 220k points in total, making the volume of the data comparable to that in \cite{b5}.
\par
More detail, we choose the dataset of 3.5 GHz to generate the UL channel response and set the number of base station antennas to 16. The dataset of 60 GHz is used to generate the DL channel response, and the number of base station antennas is set to 64. The UE is equipped with an omnidirectional
antenna. These settings are consistent with those in \cite{b5}. The BW of UL and DL channel are both set to 100 MHz with different carrier frequencies, and antenna pattern of UE and BS are set to isotropic.
\par
In the model training phase, we used the same LSTM model as in \cite{b5}.
In addition, to further improve the prediction accuracy, we combined the LSTM model with convolutional neural network, thus improving the feature extraction capability of the model. The structure and parameters of the Conv-LSTM model are shown in Fig. 5 and Table 2.
\begin{figure}[htbp]
	\centering
	\includegraphics[scale=0.18]{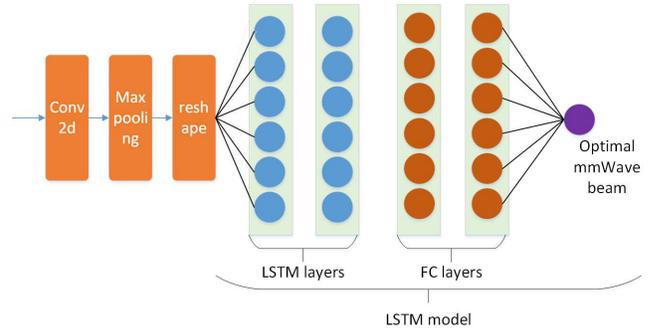}
	\caption{Convolutional layer + LSTM model.}
	\label{figure5}
\end{figure}
\begin{table}[htbp]   
	\begin{center}   
		\caption{HYPER-PARAMETERS OF THE DESIGNED MODEL}  
		\label{table:2} 
		\begin{tabular}{|c|c|}   
			\hline   \textbf{Parameter} & \textbf{Beam}  \\   
			\hline Solver & Adam  \\ 
			\hline Activation & tanh(LSTM), relu(Dense)   \\  
			\hline Batch size & 32  \\   
			\hline Max. number of epochs & 250\\
			\hline Learning rate & 0.0005\\
			\hline $L_{FC}(M_{FC})$ & 2(64,256)\\
			\hline $L_{LSTM}(M_{LSTM})$ & 2(64,128)\\
			\hline Conv2d(filters, size) & 32, 1$\times$5\\
			\hline Maxpooling(size) & 1$\times$2\\
			\hline Dataset split & 80\%-20\%\\
			\hline  
		\end{tabular}   
	\end{center}   
\end{table}
\par
The accuracy of correctly predicting the optimal beam is used as the evaluation criterion. The results of different datasets is shown in Fig. 6. The first column of the results is obtained by training the LSTM model with Wireless Insite data, and the accuracy has reached 88.20\% in \cite{b5}. As users in Wireless Insite are not moving, we assume the accuracy can reach 88.20\% at all speed. The second column of the results is obtained by training the LSTM model with DataAI-6G dataset. The accuracy of this case has reached 91.65\%, 91.03\% and 88.85\% at the speed of 72 km/h, 90 km/h and 108 km/h. Comparing these two cases, we can find that model has higher accuracy using DataAI-6G dataset, which means that our dataset provides more realistic channel features and is well adapted to the existing algorithms.
\begin{figure}[htbp]
	\centering
	\includegraphics[scale=0.65]{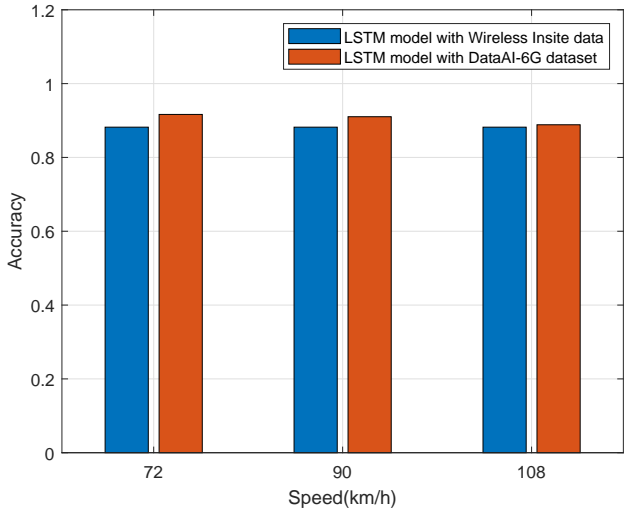}
	\caption{Beam prediction accuracy with Wireless Insite and DataAI-6G vs. Speed of movement.}
	\label{figure6}
\end{figure} 
\par
Then, with the same use of the DataAI-6G dataset comparing the accuracy of two different models, Conv-LSTM model has a better performance than LSTM model. Further observation of the transformation of accuracy with speed, as shown in Fig. 7, we can find that (i) the accuracy decreases with increasing speed, which indicates that the difficulty of beam prediction increases with speed. This is consistent with reality and reflects a high degree of realism in the mobile features of our dataset, (ii) the accuracy decreases at a more moderate rate when using Conv-LSTM model, which indicates that Conv-LSTM model has a better adaptation to speed. Therefore, borrowing algorithms from computer vision into algorithms for channel prediction is a direction that can be investigated in the future.
\begin{figure}[htbp]
	\centering
	\includegraphics[scale=0.65]{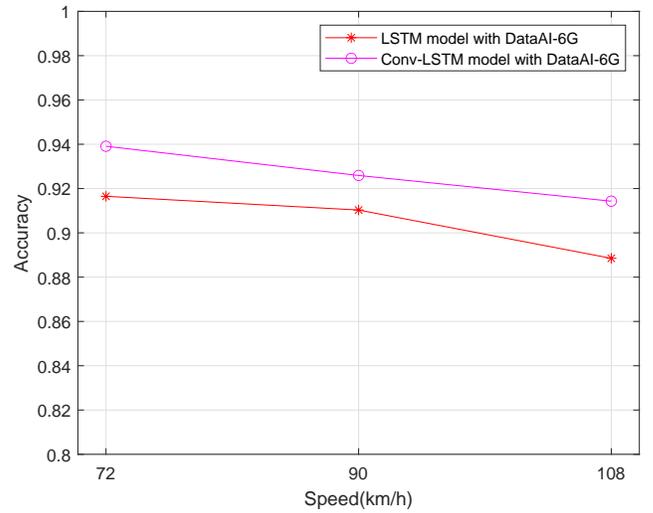}
	\caption{Beam prediction accuracy with LSTM model and Conv-LSTM model vs. Speed of movement.}
	\label{figure7}
\end{figure} 

\section{Conclusion}
Combining AI with wireless communication is a promising development direction for future 6G mobile communication.
To meet the future research, the DataAI-6G dataset takes into account the Doppler properties and, based on this, we implement a fully user-defined move function and interpolation function for the first time. Our dataset also considers spatial non-stationary properties, which are not considered in most other datasets. In the future, our dataset will further consider the communication containing RIS. And, in order to meet more research, we will also add environment and material data in future iterations of the dataset.

\end{document}